\documentclass{PoS}
\newcommand{\ba}{\begin{eqnarray}}  
\newcommand{\ea}{\end{eqnarray}}  
\newcommand{\be}{\begin{equation}}  
\newcommand{\ee}{\end{equation}}  
\newcommand{\fr}{\frac}

\newcommand{\cB}{{\cal B}}  

\newcommand{\cO}{{\cal O}}  
\newcommand{\cL}{{\cal L}}

\newcommand{\al}{\alpha}

 \newcommand{\de}{\delta} 
    
\newcommand{\s}{\sigma}

 \newcommand{\Ga}{\Gamma}

\newcommand{\im}{\mbox{Im}}  
  
\newcommand{\ugu}{\!\!\!&=&\!\!\!}

\def\tb{{\bar{t}}}
\def\sht{\hat{s}_t}
\def\shtb{\hat{s}_\tb}

\newcommand{\nn}{\nonumber} 

\newcommand{\bn}{{\bar n}}

\newcommand{\mcdot}{\!\cdot\!}

\title{The top quark jet-function at two loops}

\ShortTitle{Two-loop Jet-Function and Jet-Mass for Top Quarks}
\author{Ambar Jain\\Center for Theoretical Physics, Massachusetts Institute of
  Technology, Cambridge, MA 02139\\ E-mail:\email{ambarj@mit.edu}}
\author{\speaker{Ignazio Scimemi}
\\
        Center for Theoretical Physics, Massachusetts Institute of
  Technology, Cambridge, MA 02139\\
Departamento de F\`\i sica Te\`orica II, Universidad Complutense de Madrid, 28040 Madrid, Spain\\
        E-mail: \email{scimemi@mit.edu}}
\author{Iain W.~Stewart\\
Center for Theoretical Physics, Massachusetts Institute of
  Technology, Cambridge, MA 02139\\
  E-mail:\email{iains@mit.edu}}

\abstract{Far above threshold the production process $e^+e^-\to t\bar t$ can be
  analyzed using effective field theories. In this talk we consider the
  invariant mass distribution of top-jets and report about our computation of
  the two-loop heavy quark jet-function. This is a key part of a
  next-to-next-to-leading order analysis, and already allows for a resummation
  of all large logs which effect the shape of the top-invariant mass
  distribution at next-to-next-to-leading log order. A top-mass scheme is defined
  which is suitable for measurements involving jets, and whose anomalous
  dimension is determined by the cusp-anomalous dimension to all orders in
  perturbation theory.  }

\FullConference{8th International Symposium on Radiative Corrections (RADCOR)\\
                 October 1-5 2007\\
                 Florence, Italy}

\begin{document}

\section{Introduction}
The era of LHC is now starting and with it some new questions arise about the possibility of measuring the fundamental parameters of the  Standard Model (SM) with high precision. The issue is particularly relevant for the top quark  mass.
 The  latest Tevatron analysis give $m_t=170.9\pm 1.8$~\cite{unknown:2007bxa}, a measurement at 1\% level
  whose precision affects heavily the actual constraints on the Higgs mass as well as many new physics scenarios.
 
  
  The questions we want to address are the following: i) what observable is both
  sensitive to the mass of the top quark and under good theoretical control?
  ii) What is the theoretical framework where we can systematically describe
  this observable with high precision?  iii) How precise is our theoretical
  control over perturbative shifts to the peak of the invariant mass
  distribution?  iv) And finally, using the invariant mass for a top-mass
  measurement, what is the most appropriate (stable) mass definition to use, and
  how well can one related this mass to other existing top-mass definitions?
  The former 2 points were extensively discussed in the talk of
  A.~Hoang~\cite{Andretalk} at this meeting. The latter 2 are the focus of
  this proceedings, for which a complete discussion is given in ref.~\cite{jss}.

We concentrate on $e^+e^- \to t\bar t$ far from threshold, where
  the center-of-mass energy $Q^2\gg m_t^2$. The theoretical framework is based
  on effective field theories like HQET and SCET~\cite{SCET}, and has been outlined
  in ref.~\cite{Fleming:2007qr,Fleming:2007xt}.  Here the top quark decay
  products form well separated collinear jets together with soft-radiation among
  the jets. A thrust axis can be defined and this axis define a plane which
  divides the space in two hemispheres (that we call ``a'' and ``b''
  hemispheres).  A suitable observable is the event-shape cross-section
  $d^2\sigma/dM_t^2 dM_{\bar t}^2$. Here $M_t^2 = (\sum_{i\in a} p_i^\mu)^2$ and
  $M_{\bar t}^2 = (\sum_{i\in b} p_i^\mu )^2$ are hemisphere invariant masses.
  The different physics components of $d^2\sigma/dM_t^2 dM_{\bar t}^2$ can be
  separated by a factorization theorem derived in Ref.~\cite{Fleming:2007qr} \ba
  \label{Fthm} \fr{d\s}{d M_t^2d M_\tb^2} &= \sigma_0
  H_Q(Q,\mu_m)H_m\Big(m,\fr{Q}{m},\mu_m,\mu\Big)
 \int\! d\ell^+ d\ell^-\
 B_+\Big(\sht-\fr{Q\ell^+}{m},\Ga_t,\mu\Big) B_-\Big(\shtb-\fr{Q \ell^-}{m},\Ga_t,\mu\Big)
 \: 
\nn\\[4pt]
&\quad \times S(\ell^+,\ell^-,\mu)
 + \cO\Big(\fr{m\al_s(m)}{Q},\frac{m^2}{Q^2},\fr{\Ga_t}{m},\fr{s_t}
{m^2},\fr{s_\tb}{m^2}\Big)
 \ .
\ea
In Eq.~(\ref{Fthm}) $\sigma_0$ is the tree level Born cross section, $H_Q$ and
$H_m$ are hard-functions which encode the perturbative corrections at the scales
$Q$ and $m$, where from now on we use $m$ for the mass of the top quark. The
invariant mass variables $\hat s_t$ and $\hat s_{\bar t}$ are defined as
 $ \sht =\fr{s_t}{m}=\fr{M_t^2-m^2}{m} \,,\,
   \shtb =\fr{s_\tb}{m}=\fr{M_\tb^2-m^2}{m}\ , $
and the most sensitive region for mass measurements is the peak region where
$\hat s_{t,\bar t} \lesssim \Gamma_t + Q\Lambda_{\rm QCD}/m$. Finally, $B_{\pm}$
in Eq.~(\ref{Fthm}) are heavy-quark jet functions for the top quark/antiquark,
and $S$ is the soft function describing soft radiation between the jets. Our
main focus  here will be on the functions $B_{\pm}$, which are defined
in the heavy-quark limit $m_t \gg \Gamma_t$ using
HQET~\cite{Manohar:2000dt,Neubert:1993mb}. The soft function $S$ is universal to
massless and massive jets and a suitable model can be found in
Ref.~\cite{Hoang:2007vb}, extending earlier work in
Ref.~\cite{Korchemsky:2000kp}.


In this talk we take the first step toward next-to-next-to-leading order (NNLO)
for the invariant mass spectrum, $d^2\sigma/dM_t^2 dM_{\bar t}^2$, by computing
the top quark jet function at two-loop order.  We also give results for the
resummation of large logs for this jet function at next-to-next-to-leading log
order (NNLL). This translates into a resummation of all the large logs in the
cross-section that can modify the shape of the invariant mass
distribution~\cite{Fleming:2007qr}. We introduce a definition of the top
jet-mass scheme that has a well defined mass anomalous dimension at any order in
perturbation theory (unlike the definitions based on cutoff first moments or on peak
locations).  In this jet-mass scheme the quark-mass anomalous dimension is
completely determined by the cusp anomalous dimension at any order in
perturbation theory.

\section{ The heavy quark jet function}
\label{sec:hqet}

The jet-functions $B_\pm$ for the top quark/top anti-quark are identical by
charge conjugation, so we will only refer to the computation of $B\equiv B_+$.  $B$ is
given by the imaginary part of a forward scattering matrix element,
$ B( {\hat{s}},\delta m,\Ga_t,\mu  )=\im \big[\cB({\hat{s}},\delta m,\Ga_t,\mu) \big]\ ,$
where $\cB$ are vacuum matrix elements of a time-ordered product of fields and
Wilson lines
\begin{eqnarray} \label{eq:CB}
\cB(2 v_+\cdot r,\delta m,\Ga_t,\mu)\ugu\fr{-i}{4\pi N_c m}
 \int\! d^4x\, e^{i r\cdot x}\left\langle\,
   0\left|  T\{\bar{h}_{v_+}(0) W_{n}(0)
   W_{n}^\dagger(x)h_{v_+}(x)\}\right|0\right\rangle 
 \,.
\end{eqnarray}
Here $v_+^\mu$ is the velocity of the heavy top quark, and we introduce
null-vectors $n^\mu$ and $\bn^\mu$ so that we can decompose momenta as $p^\mu=
n^\mu \bn\mcdot p/2 + \bn^\mu n\mcdot p/2 + p_\perp^\mu$.  The vectors satisfy
$v_+^2=1$ and $n^2=\bn^2=0$, and the definition of the Wilson lines in
Eq.~(\ref{eq:CB}) is
\begin{eqnarray} \label{WW}
 W^\dagger_{n}(x) &=  {\rm P}\exp\Big(ig\int_0^\infty ds\, \bar{n}\cdot A_n(\bar
 n s+x)\Big),\quad
  W_{n}(x) =\overline {\rm P}\exp\Big(-ig\int_0^\infty ds\, 
  \bar{n}\cdot A_n(\bar n s+x)\Big) \,.
\end{eqnarray}
These Wilson lines make ${\cal B}$ gauge-invariant and encode the residual
interactions from the antitop jet.  The HQET fields $h_{v_+}$ have the leading
order Lagrangian
\begin{eqnarray} \label{eq:LHQET}
 \cL_h &= \bar{h}_{v_+} \Big(iv_+\cdot D-\de
 m+\fr{i}{2}\Ga_t\Big)h_{v_+} \,.
\end{eqnarray}
Here $\Ga_t$ is the top quark total width, obtained from matching the top-decay
amplitudes in the standard model (or a new physics model) onto HQET at leading
order in the electroweak interactions, and at any order in $\al_s$.
The residual mass term
$\de m$ in Eq.~(\ref{eq:LHQET}) fixes the definition of the top mass $m$ for the
HQET computations~\cite{Falk:1992fm}, where
 $\de m= m_{pole}-m \,.$
 From the definitions in Eq.~(\ref{eq:CB}) and the Lagrangian in
 Eq.~(\ref{eq:LHQET}) one can deduce a series of properties of the jet function,
 which state that it is easy to reconstruct $B(\hat s,\delta m,\Gamma_t,\mu)$
 from $B(\hat s,0,0,\mu)$. In particular in ${\cal B}$ the $\hat s$, $\delta m$,
 and $\Gamma_t$ dependence formally occurs only in the combination $(\hat
 s-2\delta m + i\Gamma_t)$. For this reason it is useful to have a notation for
 computations done with a zero residual mass term in the Lagrangian, and with
 zero-width. Thus we define 
\ba B(\hat{s},\delta m,\mu) \equiv B(\hat{s},\delta
 m, 0,\mu) \,, \qquad 
  & {\cal B}(\hat s,\delta m,\mu) &\equiv {\cal B}(\hat s,\delta
 m,0,\mu) \,,
 \nn\\
 B(\hat s,\mu) \equiv B(\hat s,0,0,\mu) \,, 
\qquad 
& {\cal B}(\hat s,\mu) & \equiv
 {\cal B}(\hat s,0,0,\mu) \,.  \ea


\section{Non-Abelian exponentiation and  jet-mass}
 
It is possible to rewrite Eq.~(\ref{eq:CB}) as a matrix element of pure Wilson
lines, 
\ba \label{cBwloop} \cB(2 v\cdot r,\mu) 
&= \fr{i}{2\pi N_c m} \int\! dx^0\, e^{i v\cdot r\, x^0}\: \theta(x^0)\,
\left\langle\, 0\left| {\rm tr}\: T\: W_v^\dagger(0) W_{n}(0) W_{n}^\dagger(x)
    W_v(x) \right|0\right\rangle
 \,,
\ea
where $2v\cdot r=\hat s$, we use the shorthand $x^0=v\cdot x$, and the trace ${\rm
  tr}$ is over color indices.
The definition of $W_v$ and $W_v^\dagger$ is as in Eq.~(\ref{WW}) with $n\to v$.
Following the steps outlined in the original paper,~\cite{jss}, also the Fourier
transform of the jet function comes out as a product of Wilson lines, 
\ba
\label{Bwloop} B(\hat s,\mu) = \frac{1}{2\pi} \int \! dy \; e^{i \hat s\,y} \;
\tilde B(y,\mu) \,, && \ \ \tilde B(y,\mu) = \frac{1}{ m\, N_c}\: \big\langle 0
\big| {\rm tr} \,\big[\, \overline T\, W_n^\dagger(2y) W_v(2y) \big] \big[\, T\,
W_v^\dagger(0) W_n(0) \big] \big| 0 \big\rangle \,.  \nn \\
\ea where $y=y-i0$ to ensure convergence as $\hat s\to \infty$.  Due to the
non-abelian exponentiation theorem~\cite{Gatheral:1983cz,Frenkel:1984pz},
$\tilde B(y,\mu)$ exponentiates.
Thus we can write the result of our two-loop computation as~\cite{jss}:
\ba
\label{Bnonabelian} m \tilde B(y,\mu) & = \exp\Bigg\{
\frac{C_F\alpha_s(\mu)}{\pi} \bigg( \tilde L^2 + \tilde L+\frac{\pi
  ^2}{24}+1\bigg) + \frac{\alpha_s^2(\mu) C_F \beta_0}{\pi^2} \bigg[ \frac{1}{6}
\tilde L^3 +\frac{2}{3}\tilde L^2 +\frac{47}{36}\tilde L -\frac{\zeta
  (3)}{48}+\frac{5 \pi
  ^2}{576}+\frac{281}{216}\bigg] \nn\\
&\quad \!\!\!  + \frac{\alpha_s^2(\mu) C_F C_A}{\pi^2} \bigg[
\Big(\frac{1}{3}-\frac{\pi ^2}{12}\Big) \tilde L^2 +
\bigg(\frac{5}{18}-\frac{\pi ^2}{12}-\frac{5 \zeta_3}{4}\bigg) \tilde L
-\frac{5\zeta_3}{8}-\frac{17 \pi ^4}{2880}+\frac{7 \pi
  ^2}{144}-\frac{11}{54}\bigg] \Bigg\} \,, 
\ea
where $ \tilde L \, \equiv \, \ln\big( i e^{\gamma_E} y\,\mu \big) \,.  $ The
non-abelian exponentiation theorem guarantees that corrections to this result
are ${\cal O}(\alpha_s^3)$ in the exponent, and that these corrections vanish if
we take the abelian limit $C_A\to 0$ and $n_f\to 0$. Since the exponent of the
abelian result is one-loop exact, we can use it to test the perturbative
behavior of different definitions of the top-mass at any desired order in
pertubation theory.  Choosing an appropriate top-mass definition corresponds to
choosing an appropriate $\delta m$. In ref.\cite{jss} we explored several
possibilities and came to the following definition, which we refer to as the
jet-mass scheme
 \ba \label{mscheme}
 && \delta m_J = \frac{-i}{2\,  \tilde {B}(y,\mu)}\: 
   \frac{d}{dy}\, \tilde {B}(y,\mu) \bigg|_{y = -ie^{-\gamma_E}/R}
   = e^{\gamma_E}\: \frac{R}{2}\, \frac{d}{d\ln(i y)} \ln \tilde {B}(y,\mu) 
  \bigg|_{iye^{\gamma_E} =1/R}
    \,. \nn
\ea
The scheme depends on a parameter $R$, and we must take $R\sim \Gamma_t$ in
order to satisfy the power counting criteria.  Different choices for $R$ specify
different schemes, and are analogous to the difference between the ${\rm MS}$
and $\overline {\rm MS}$ mass-schemes.  The scheme in Eq.~(\ref{mscheme}) is
free from leading renormalon ambiguities~\cite{FHMSmass:inprep}.  Let us now
check that the mass so defined has good transitivity properties.  Transitivity
is a well-known feature of the $\overline {\rm MS}$ mass, and implies that we
will obtain the same result if we evolve directly from $\mu_0\to \mu_2$, or if
we first evolve from $\mu_0\to \mu_1$ and then from $\mu_1\to \mu_2$.
Transitivity is guaranteed by any mass-scheme with a consistent anomalous
dimension and renormalization group equation.  Since in HQET the scale
independent $m^{\rm pole}= m(\mu) +\delta m(\mu)$, the general form for the RGE
equation for the mass is
\ba \label{mad}
  \mu\frac{d}{d\mu} m(\mu) &= \gamma_m[R, m(\mu),\alpha_s(\mu)] \,,
\qquad\qquad
  \gamma_m =  - \mu\frac{d}{d\mu} \delta m(\mu) \,,
\ea
where $R$ is a mass dimension-1 scheme parameter. 
To all orders in perturbation theory, using Eq.~(\ref{mscheme}), the
jet-mass anomalous dimension is~\cite{jss}
\ba
  \gamma_m^J = - \frac{d\delta m(\mu)}{d\ln\mu} = - e^{\gamma_E} \frac{R}{2}\,
  \frac{d}{d\ln\mu}\, \frac{d}{d\ln(i y)}\,  \ln \tilde B(y,\mu)
  \bigg|_{i ye^{\gamma_E}=1/R}= - e^{\gamma_E} R\ \Gamma^{\rm c}[\alpha_s(\mu)]\,.
\ea 
Thus, to all
orders in perturbation theory the jet-mass scheme, has a consistent anomalous dimension as in Eq.~(\ref{mad}),
and yields a transitive running mass, $m_J(\mu)$. The final anomalous dimension
equation for the jet-mass is
fully determined by the cusp-anomalous dimension $\Gamma^{\rm c}$, which is
known to three-loop order~\cite{Moch:2004pa}.
Note that the form of the anomalous dimension in $\mu d/d\mu\,[ m_J(\mu)/R]$ has
the same structure as that in $\mu d/d\mu\,[ \ln \overline{m}(\mu)]$, where
$\overline{m}(\mu)$ is the $\overline {\rm MS}$ mass. 
\section{Results for the NNLL Jet Function }
\begin{figure}[t!]
  \centerline{ 
   \includegraphics[width=7cm]{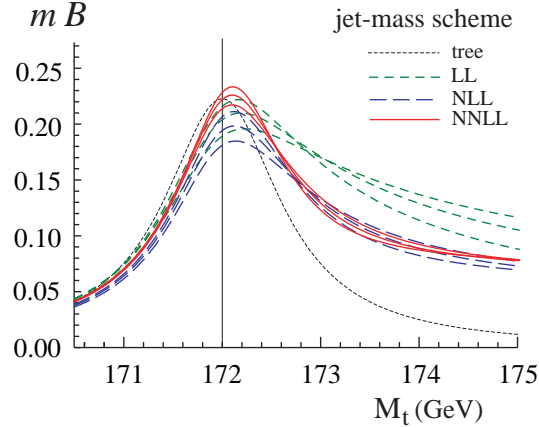}
  }
\vskip-0.2cm
\caption{
  The jet function, $m B(\hat s,\delta m_J,\Gamma_t,\mu_\Lambda,\mu_\Gamma)$
  versus $M_t$, where $\hat s=(M_t^2-m^2)/m$ and $\Gamma_t=1.43\,{\rm GeV}$, and
  $R= 0.8\,{\rm GeV}$. The black dotted curve is the tree-level Breit-Wigner,
  the green short-dashed curves are LL results, blue long-dashed curves are NLL,
  and the solid red curves are at NNLL order. For each order we show three
  curves with $\mu_\Gamma=3.3,5.0,7.5\,{\rm GeV}$ respectively.  Other parameter
  choices are discussed in \cite{jss}.  }
\label{fig:BSrun}
\end{figure}

In this section we discuss the final result for the heavy quark jet function
$B(\hat s,\delta m,\Gamma_t,\mu)$, with NNLO perturbative corrections and a NNLL
resummation of large logs. We have studied the numerical effect of these
two-loop corrections as well as of the log-resummation, including the
perturbative convergence and $\mu$-dependence of $B$ as a function of $\hat s$.
At tree-level $B(\hat s,\delta m,\mu)=\delta(\hat s)$ and $B(\hat s,\delta
m,\Gamma_t,\mu)$ is simply a Breit-Wigner centered at $\hat s=0$ with a width
$\Gamma_t$. Beyond tree-level the jet function becomes dependent on $\mu$ and on
the choice of mass-scheme through $\delta m$.  For the cross-section
$d^2\sigma/dM_t^2 dM_{\bar t}^2$ in Eq.~(\ref{Fthm}) it has been proved that at
any order in perturbation theory, the only large logs that effect the shape of
the invariant mass distribution are those due to the resummation in the
heavy-quark jet function~\cite{Fleming:2007xt}.  Furthermore these large logs
only exist between scales $\mu_\Gamma\sim \Gamma\equiv \Gamma_t+ Q\Lambda_{\rm
  QCD}/m$ and $\mu_\Lambda \gtrsim \Lambda_{\rm QCD}+ m\Gamma_t/Q$.  The remaining large logs
only modify the cross-section normalization. The expression which sums all logs
between the scales $\mu_Q\simeq Q\gg \mu_m\simeq m\gg \mu_\Gamma \simeq
\Gamma\gg \mu_\Lambda\gtrsim \Lambda_{\rm QCD}$ is \ba
\label{Fthm2} \frac{d^2\sigma}{dM_t dM_{\bar t} } \ugu 4 \sigma_0 M_t M_{\bar
  t}\, H_Q(Q,\mu_Q) U_{H_Q}(Q,\mu_Q,\mu_m) H_m(m_J,\mu_m)
U_{H_m}(Q/m_J,\mu_m,\mu_\Lambda)
\\
& & \hspace{-2cm} \times \int_{-\infty}^{+\infty}\!\!\!\!\!\! d\ell^+\,d\ell^-
B_+\Big(\hat s_t -\frac{Q\ell^+}{m_J} ,\delta
m_J,\Gamma_t,\mu_\Lambda,\mu_\Gamma\Big) B_-\Big(\hat s_{\bar t}
-\frac{Q\ell^-}{m_J},\delta m_J,\Gamma_t,\mu_\Lambda,\mu_\Gamma\Big)
S\big(\ell^+,\ell^-,\mu_\Lambda,\delta,\bar\Delta(\mu_\Lambda)\big) \,,\nn \ea
where we have defined the resummed jet function as \ba \label{eq:bmumu} B(\hat s
, \delta m_J,\Gamma_t, \mu_\Lambda,\mu_\Gamma) &\equiv \int\!\! d\hat s'\ 
U_B(\hat s-\hat s',\mu_\Lambda,\mu_\Gamma) \ B(\hat s',\delta
m_J,\Gamma_t,\mu_\Gamma)\,. \ea 
Since the scales $\mu_\Gamma$ and $\mu_\Lambda$ differ by a factor of $Q/m \gg
1$ it is necessary to sum the large logs between these scales.  In
Eqs.~(\ref{Fthm2},\ref{eq:bmumu}) large logs are resummed by the evolution
factors $U_{H_Q}$, $U_{H_m}$, and $U_B$, and of these, the first two only affect
the overall normalization.  The numerical importance of the resummation of all
large logs was demonstrated at NLL order in Ref.~\cite{Fleming:2007xt}.

In fig.\ref{fig:BSrun} we the plot LL, NLL, and NNLL results for the jet
function. We observe that the jet-mass scheme results exhibit good perturbative
convergence with a stable peak location for the jet-function.
In the jet-mass scheme the scale dependence in the slope before the peak is
$\sim 6\%$ at NLL and $\sim 2\%$ at NNLL, while the maximum variation near the
peak is $14\%$ at NLL and $7\%$ at NNLL, and then in the tail above the peak it
is $\sim 12\%$ at NLL and $\sim 5\%$ at NNLL. Thus, in the jet-mass scheme the
$\mu_\Gamma$ dependence is reduced by a factor of two or more. The same level of
improvement is also observed for different mass-scheme parameters $R$ than the
one shown.

\section*{Acknowledgments}
This work was
supported in part by the Department of Energy Office of Nuclear Science under
the grant DE-FG02-94ER40818.  I.~Scimemi was also supported in part by a Marie
Curie International Fellowship from the European Union, grant number 021379
(BDECMIT), and thanks the Fundaci\'o Bosch i Gimpera of the
University of Barcelona (Spain) for support.  I.W.~Stewart was also supported in
part by the DOE Outstanding Junior Investigator program and Sloan Foundation.

\label{sect:results}

\end{document}